\documentstyle[prl,aps,epsf,twocolumn]{revtex}
\newcommand{\eq}{\begin{equation}}
\newcommand{\ee}{\end{equation}}
\newcommand{\s}{{\sigma}}

\newcommand{\vq}{{\vec{q}}}

\def\a{\alpha}

\def\e{\epsilon}
\def\o{{\omega}}

\def\half{{1\over2}}

\def\intq{{\int {d^2q\over(2\pi)^2}}}

\def\ua{\uparrow}
\def\da{\downarrow}
\def\eqa{\begin{eqnarray}}
\def\eea{\end{eqnarray}}
\def\prl{{Phys. Rev. Lett.}}
\def\prb{{Phys. Rev. {\bf B}}}

\def\jpcm{{Jour. Phys. Cond. Matt.}}
\def\jpsj{{Jour. Phys. Soc. Japan\ }}

\def\rmp{{Rev. Mod. Phys.}}
\def\ssc{{Sol. State Comm.}}
\begin{document}
\draft
\flushbottom
\twocolumn[
\hsize\textwidth\columnwidth\hsize\csname @twocolumnfalse\endcsname
\title{The Effects of Disorder on the $\nu=1$ Quantum 
Hall State}
\author{  Ganpathy Murthy}
\address{
Department of Physics and Astronomy, University of Kentucky,
Lexington, KY 40506}
\date{\today}
\maketitle
\tightenlines
\widetext
\advance\leftskip by 57pt
\advance\rightskip by 57pt

\begin{abstract}
A disorder-averaged Hartree-Fock treatment is used to compute the
density of single particle states for quantum Hall systems at filling
factor $\nu=1$. It is found that transport and spin polarization
experiments can be simultaneously explained by a model of mostly
short-range effective disorder. The slope of the transport gap (due to
quasiparticles) in parallel field emerges as a result of the interplay
between disorder-induced broadening and exchange, and has implications
for skyrmion localization.

\end{abstract}
\vskip 1cm

]
\narrowtext
\tightenlines
\section{Introduction}
\label{intro}
The discovery of the quantum Hall effects\cite{iqhe-ex,fqhe-ex} has
led to a new appreciation of the possible ground states of the
two-dimensional electron gas (2DEG) in high magnetic fields. In a high
field $B$, the kinetic energy is quantized into Landau levels
(LLs) with energy $(n+\half)\o_c$, where $\o_c=eB/m$ is the cyclotron
frequency. Each LL has a macroscopic degeneracy of $BA/\phi_0$, where
$A$ is the area of the 2DEG, and $\phi_0=h/e$ is the quantum unit of
flux. 

The integer quantum Hall effect (IQHE) can be understood at a
single-particle level after including the effects of
disorder\cite{perspectives}. Interactions are also very important at
$\nu=1$. It is known that the $\nu=1$ state is a spontaneous
ferromagnet with charged skyrmionic excitations in the clean
limit\cite{shivaji-skyrmion}, which have been seen in
experiments\cite{skyrmion-ex,goldberg}. A complete understanding of
the $\nu=1$ state should involve the effects of both interactions and
disorder.

One of the physical properties that points to a role for disorder is
the small magnitude of the transport gap for
$\nu=1$\cite{nicholas,schmeller}. Even after assuming that skyrmions,
which have smaller excitation energy than
quasiparticles\cite{shivaji-skyrmion,fertig}, are the charge carriers,
and that sample thickness and Landau level
mixing\cite{cooper,bonesteel,mihalek} contribute to reducing this
excitation energy further, the predicted gap is nearly a factor of two
above the data\cite{nicholas,schmeller}. In this paper we will take a
straightforward phenomenological approach to computing the physical
properties of the $\nu=1$ state in the presence of both disorder and
interactions\cite{ando-uemura}. In this approach the single-particle
Green's function in the Hartree-Fock approximation is averaged over
disorder to obtain a Dyson equation
\eqa
\bigg({\cal{G}}\bigg)_{nn}^{-1}=&\bigg({\cal{G}}_0\bigg)_{nn}^{-1}-
\intq U(q) \sum\limits_{m} |\rho_{nm}(\vq)|^2 {\cal{G}}_{mm}\\
=&\bigg({\cal{G}}_0\bigg)_{nn}^{-1}-E_C^2
\sum\limits_{m} \a_{mn} {\cal{G}}_{mm}
\label{dyson}
\eea
where ${\cal{G}}_0$ is the Green's function in the clean limit seen as
a matrix in the the LL indices, $\cal{G}$ is the disorder-averaged
Green's function, $\rho_{nm}(\vq)$ is the matrix element of the
electron density operator $\hat{\rho}(\vq)$ between the LLs $n$ and
$m$, $E_C=e^2/\varepsilon l_0$, and 
\eq
<V_r(\vq)V_r(\vq')>=U(q)(2\pi)^2\delta^2(\vq+\vq')
\ee
defines $U(q)$ in terms of the ensemble average of the disorder
potential $V_r$. The interaction enters the picture through the
 HF energies which enter ${\cal G}_0$
\eq
\e_{0\s}(n)=n\o_c- {\s E_Z\over2}-\int_q v(q) \sum\limits_{m}|\rho_{nm}(\vq)|^2 N_{F\s}(m)\nonumber
\ee
where $\s=\pm1$ is the spin index, $E_Z=g^*\mu B_{tot}$ is the Zeeman
coupling, $v(q)$ is the electronic interaction, and $N_{F\s}(m)$ is
the Fermi occupation of the (spin-split) LL with index $m$.  This set
of equations can be self-consistently solved to obtain the chemical
potential, the density of states (given by the imaginary part of the
Green's function) and occupations of various Landau
levels. Subsequently one can calculate any physical quantity that
depends only on single-particle properties.

We will be concerned with high-mobility samples, where the 2DEG is
separated from the dopant atoms by an insulating layer of thickness
$d\approx 1000\AA\gg l_0$  (the magnetic length). The
disorder potential arises from the Coulomb interaction of the
electrons with the fluctuations in the density of the dopant atoms. In
an incompressible state, one might naively assume that this potential
is not screened. However, this leads to divergent fluctuations on
large length scales\cite{gergel}. A number of approaches have been
proposed to handle the screening
problem\cite{screening-ando,screening-gedhardts,screening-dassarma}. In
the following analysis we will be guided by the approach of Efros and
coworkers\cite{efros}, in which the bare disorder potential is
screened nonlinearly. This leads to a picture of the 2DEG divided into
incompressible strips where the density is more or less pinned to an
integer value, and more compressible ``metallic'' regions where
quasiparticles and quasiholes nucleate to screen the bare disorder
potential, and the density lies between two integer
values\cite{efros}. In the incompressible strips the long-range
potential still makes its presence felt, but is screened by the
``metallic'' regions beyond the width of the incompressible
strips\cite{efros} $W\gg l_0$. This picture has recently been supported by
scanning probe experiments which image the electron density of the
2DEG\cite{imaging}. The images show that even within the
incompressible strips there are short-range density fluctuations.
Based on this picture, we will assume an effective disorder potential
which is mostly short-ranged, but also has a small long-range
component which is screened\cite{efros} for $q\le W^{-1}$. In this
case, due to the small-$q$ properties of the density matrix elements
$\rho_{mn}$, the long-range part of the effective disorder adds a
constant to all the diagonal elements $\a_{mm}$ and makes a negligible
contribution to the off-diagonal terms. We model the short-range part
of the disorder with uncorrelated $\delta$-function impurities, with
$U(q)=\a_sE_C^2l_0^2$, where $\a_s$ is the disorder strength. Our
model disorder potential is characterized by the two dimensionless
parameters $\a_l$ and $\a_s$ and has the simple form
$\a_{mn}=\a_s+\a_l\delta_{mn}$.

Consider the effect of disorder on the spin polarization and the
transport gap. In the clean limit, the $n=0,\ua$ LL is completely
occupied while the $n=0,\da$ and all other LLs are empty. The
transport gap is just the splitting between the $n=0,\ua$ and
$n=0,\da$ LLs. Under realistic conditions (magnetic fields of a few
Tesla) the interaction energy $E_C$ is usually larger than the
cyclotron energy, and both the above scales are much larger than the
Zeeman coupling $E_Z$. At $\nu=1$ the exchange energy dominates the
gap. The splitting between the $n=0,\ua$ and $n=0,\da$ levels (assuming
that only these two are occupied) is
\eq
\Delta=E_0(N_F(\ua,0)-N_F(\da,0))+E_Z
\label{exch-gap}\ee
where the exchange integral is $
E_0=\intq v(q) e^{-q^2l_0^2/2}$.

Eq(\ref{exch-gap}) shows that disorder can be expected to reduce the
gap\cite{ando-uemura}. If the disorder broadening of the LLs is
sufficient to make the single-particle states in $n=0,\ua$ and
$n=0,\da$ overlap, then the gap will be reduced when compared to the
clean limit according to Eq(\ref{exch-gap}). However, the band overlap
will simultaneously decrease the spin polarization. In
order to see if this effect is operative in real samples,  one
needs to  look at the spin polarization.

Aifer {\it et al}\cite{goldberg} made measurements of the absolute
spin polarization near $\nu=1$ using an optical absorption
technique. Their data show a ``flat top'', demonstrating that for
$0.95\le\nu\le1.05$ the sample is {\it only 65\% spin polarized even
at the lowest temperatures}, with an estimated error of
10\%\cite{bennett}. A similar feature has recently been observed in
optically pumped nuclear magnetic resonance (OPNMR)
measurements\cite{khandelwal}.  Outside this range of $\nu$ the
data show unambiguous evidence for
skyrmion-induced depolarization\cite{skyrmion-ex,goldberg,khandelwal}.
Figure 1 shows our predictions for the spin polarization near $\nu=1$
for short range + small amounts of long-range disorder.  A Fang-Howard
form\cite{fang-howard} for the interaction $v(q)$ with width
$b^{-1}=0.5l_0$ was used to compute exchange integrals, the $n=0,1,2$
Landau levels were kept, and other parameters appropriate to Sample A
of Aifer {\it et al}\cite{goldberg} were used ($B_{\nu=1}=6.2T$).

\begin{figure}
\narrowtext
\epsfxsize=2.4in\epsfysize=2.4in
\hskip 0.3in\epsfbox{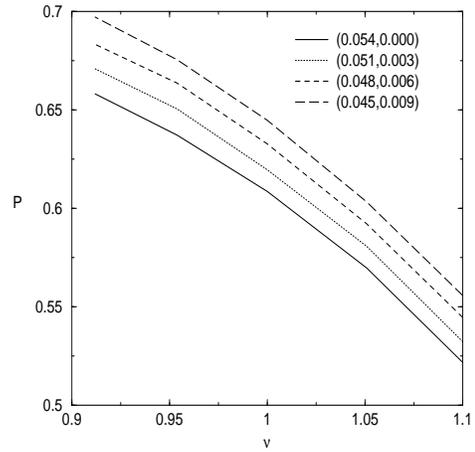}
\vskip 0.15in
\caption{Spin polarizations in a small range around $\nu=1$ for different 
combinations of short and long range disorder. The legend shows the values 
$(\a_s,\a_l)$. The polarization is seen to be robust. 
\label{fig1}}
\end{figure}
The lack of full polarization\cite{goldberg} shows that the
disorder-broadened $n=0,\ua$ and $n=0,\da$ LLs {\it do} overlap,
so by Equation (\ref{exch-gap}) we can expect reductions of the
transport gap\cite{nicholas,schmeller}. For definiteness, we will
focus on the data of Schmeller {\it et al}\cite{schmeller}, who
measured the transport gap $\Delta$ at $\nu=1$ as a function of the
Zeeman coupling $E_Z$.  There were two noteworthy features in their
data. Firstly, the largest gap they obtained (in
the SI1 sample) for $\nu=1$ (at $E_Z=0.01E_C$) was approximately
$0.25E_C$, almost a factor of two smaller than the smallest
theoretical estimate for the transport
gap\cite{cooper,bonesteel,mihalek}. Secondly, they
observed a high slope of $\Delta$ at small $E_Z$, and interpreted the
result as showing evidence for large skyrmions. If skyrmions with $s$
reversed spins are the charge carriers, then the transport gap should
behave as
\eq
\Delta(E_Z)=\Delta_0+sE_Z
\label{skyrmion-slope}\ee
Thus interpreted, the data suggest $s\approx7$ skyrmions.

Figure 2 shows the results of our self-consistent calculations of the
transport gap. Once again the Fang-Howard form for the interaction
with width $b^{-1}=0.5l_0$ was used, with the parameters appropriate
to the SI1 sample ($B_{\perp}=2.3T$) of Schmeller {\it et
al}\cite{schmeller}. Since the zero-field mobilities of the SI1 sample
($\approx 3.4\times 10^6 cm^2/Vs$)\cite{schmeller} and Sample A of
Aifer {\it et al} ($\approx 3.2\times 10^6 cm^2/Vs$)\cite{goldberg}
are similar, we can roughly expect the same disorder strengths in the
two samples. Based on this expectation, some of the same disorder
strengths as in Figure 1 have been used in Figure 2. We assume that
the extended quasiparticle states of the $n=0,\ua$ and $n=0,\da$ LLs
lie at the respective band centers, defined as the energy where the
density of states of a particular band is the maximum (this is exact
when there is no LL mixing due to disorder\cite{liu}). To
conform to convention, based on fitting to the form $R_{xx}\simeq
e^{-\Delta/2T}$, the transport gap is computed as twice the difference
between the chemical potential and the nearest extended state.
\begin{figure}
\narrowtext
\epsfxsize=2.4in\epsfysize=2.4in
\hskip 0.3in\epsfbox{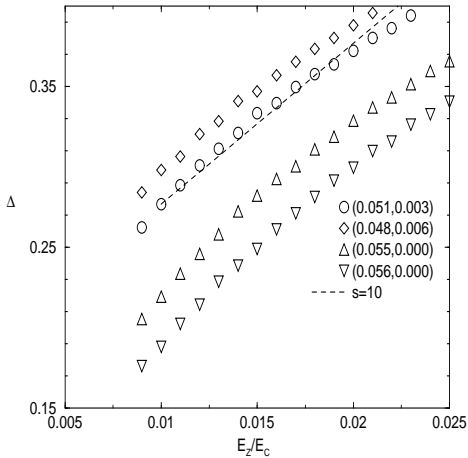}
\vskip 0.15in
\caption{Transport gaps (in units of $E_C$) as a function of $E_Z/E_C$ 
for some of the same combinations of disorder as in Figure 1. Once
again the legend indicates the values of $(\a_s,\a_l)$. The dashed
line with slope $s=10$ is a guide to the eye.
\label{fig2}}
\end{figure}
As can be seen, every combination of disorder considered produces a
high slope of $\Delta$ vs $E_Z$, and the slope is almost independent
of the magnitude of the gap in this region of $E_Z$. Similar results
are obtained for other realistic values of sample thickness
$b^{-1}$. If interpreted according to Eq(\ref{skyrmion-slope}) this
would correspond to $s\approx10$ skyrmions. The implication here is
that high slope seen in the data could be the result of disorder +
exchange, rather than skyrmions.
In fact, such an explanation was proposed in
earlier experimental work\cite{nicholas} which also saw high slopes of
$\Delta$ versus $E_Z$ ($s\approx6-20$). Nicholas {\it et al} used an
empirical gaussian form for the DOS\cite{nicholas}, while here we
obtain it from a self-consistent treatment of disorder and
interactions.

\begin{figure}
\narrowtext
\epsfxsize=2.4in\epsfysize=2.4in
\hskip 0.3in\epsfbox{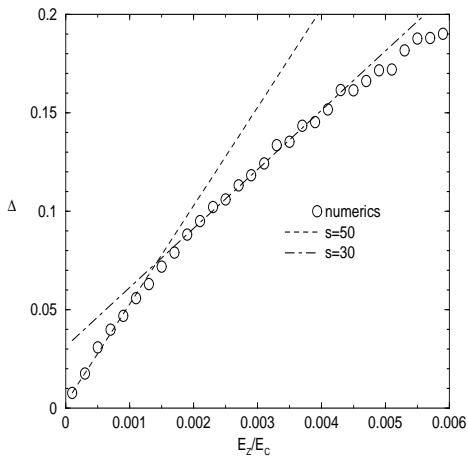}
\vskip 0.15in
\caption{Transport gap (in units of $E_C$) as a function of $E_Z/E_C$
for a sample near the spin-collapse transition. Very high slopes,
corresponding to $s=25-50$ can be seen in different parts of the
numerical results. The dashed and dash-dotted lines are guides to the
eye. We have used $\a_s=0.054,\ \a_l=0$.
\label{fig3}}
\end{figure}
Recently there have been reports of very large
skyrmions\cite{large-skyrmion,skyrmion-search} ($s$ in the range
$36-50$) for systems with very small effective $E_Z$.  
 This can also be
understood within the context of our model. It was shown by Fogler and
Shklovskii\cite{spin-collapse} that as disorder increases beyond a
critical value the exchange gap collapses (in the absence of Zeeman
coupling). Near the critical value of disorder, even a small change in
$E_Z$ makes a tremendous difference in the transport gap. Figure 3
shows a plot of the transport gap as a function of $E_Z$. If the
initial slope is interpreted in terms of Equation
(\ref{skyrmion-slope}) it would correspond to very large skyrmions
($s=50$). The slope decreases with increasing $E_Z$, as in the
data\cite{large-skyrmion,skyrmion-search}.

Despite the semi-quantitative agreement with many experiments at
$\nu=1$, there are several caveats that should be mentioned. We have
ignored correlation effects beyond Hartree-Fock. In addition to
important qualitative phenomena such as the coulomb
gap\cite{coul-gap}, these effects can also decrease the transport
gap\cite{correlation-correction}.  In the experimentally relevant
range of sample densities these effects are estimated to be
unimportant for $\nu=1$, but are significant for $\nu=3$ and
higher\cite{correlation-correction}. Secondly, the parallel field
causes mixing between the different electric subbands, leading to
modifications of the interaction\cite{parallel-field}. These effects
are also expected to be more significant for higher LLs. Finally, we
have assumed that the extended states lie at the band
maxima. Numerical work on noninteracting models with disorder shows
that this is a good approximation if LL mixing due to
disorder is weak, but fails for large LL mixing\cite{liu}. This effect
is also more significant at higher $\nu$.  In this context, the
transport gaps for $\nu=3$ have also been measured as a function of
$E_Z$. A straightforward application of the methods of this paper
gives a high slope similar to $\nu=1$, in agreement with earlier work
(see Usher {\it et al}\cite{nicholas}) but disagrees with other
data\cite{schmeller}. It is possible that the disorder + exchange
mechanism does occur, but is modified in a sample-dependent way by the
other effects mentioned
above\cite{correlation-correction,parallel-field,liu}, which are all
expected to be substantially greater for $\nu=3$ than for
$\nu=1$. Work is in progress to take these effects into
account\cite{future} with a numerical HF approach for concrete
disorder realizations\cite{numerical-hf}.

If one assumes that the disorder + exchange mechanism is the dominant
one for $\nu=1$, there are some interesting consequences. The slope of
the transport gap as a function of $E_Z$ is consistent with skyrmions
being invisible in transport at $\nu=1$. However, from polarization
measurements\cite{skyrmion-ex,goldberg,khandelwal} we know that
skyrmions depolarize the system away from $\nu=1$. This means that
skyrmions, if they do exist at $\nu=1$, must be localized. This is
plausible, since skyrmions are extended objects, and therefore their
``hopping'' requires many spin overlaps. These overlaps are presumably
reduced by disorder, possibly leading to skyrmion localization. In
fact, the most recent OPNMR measurements\cite{khandelwal} are
consistent with skyrmion localization in a small range around $\nu=1$,
though the analysis is complicated by wavefunction and polarization
profiles perpendicular to the 2DEG\cite{sinova}.

In summary, the interplay between disorder and
exchange\cite{ando-uemura}, when treated in a self-consistent
disorder-averaged Hartree-Fock approximation, can provide a consistent
account of many of the experimental observations at
$\nu=1$\cite{goldberg,nicholas,schmeller,large-skyrmion,skyrmion-search}. This
approach offers an alternative interpretation of the transport
data\cite{nicholas,schmeller,large-skyrmion,skyrmion-search} which
does not include skyrmions, and suggests that skyrmions may be
localized\cite{khandelwal} and hence invisible in transport.  In order
to test the model rigorously, it would be very helpful to have
measurements of transport gaps and spin polarization on the same
sample at $\nu=1$.  More work is needed at higher $\nu$ to
tease apart the confounding effects of Landau level mixing due to
disorder, correlation effects, and effects due to the parallel field.

It is a pleasure to thank J. P. Eisenstein, H. A. Fertig,
B. B. Goldberg, A. H. MacDonald, and especially S. Das Sarma for many
illuminating conversations. The author also wishes to thank the
National Science Foundation for partial support (under DMR 0071611).

\end{document}